\documentclass[10pt,aps,prb,twocolumn,showpacs,amssymb]{revtex4-1}
\usepackage{amsmath,graphics,epsfig,mathrsfs}
\usepackage{bm}
\usepackage{color}
\usepackage{hyperref}
\hypersetup{backref=true,pagebackref=true,hyperindex=true,colorlinks=true,breaklinks=true,urlcolor=
red,linkcolor=blue,bookmarks=true,bookmarksopen=true,citecolor=red}

\begin{document}
\title{Robust spin transfer torque in antiferromagnetic tunnel junctions}
\author{Hamed Ben Mohamed Saidaoui}
\affiliation{Physical Science and Engineering Division, King Abdullah University of Science and Technology (KAUST), Thuwal 23955-6900, Kingdom of Saudi Arabia}
\author{Xavier Waintal}
\affiliation{CEA-INAC/UJF Grenoble 1, SPSMS UMR-E 9001, Grenoble F-38054, France}
\author{Aurelien Manchon}
\affiliation{Physical Science and Engineering Division, King Abdullah University of Science and Technology (KAUST), Thuwal 23955-6900, Kingdom of Saudi Arabia}

\begin{abstract}
We theoretically study the current-induced spin torque in antiferromagnetic tunnel junctions, composed of two semi-infinite antiferromagnetic layers
separated by a tunnel barrier, in both clean and disordered regimes. We find that the torque enabling the electrical manipulation of the N\'eel antiferromagnetic order parameter is out of plane $\sim {\bf n}\times{\bf p}$, while the torque competing with the antiferromagnetic exchange is in-plane $\sim {\bf n}\times({\bf p}\times{\bf n})$. Here, ${\bf { p}}$ and ${\bf { n}}$ are the N\'eel order parameter direction of the reference and free layers, respectively. Their bias dependence shows similar behavior as in ferromagnetic tunnel junctions, the in-plane torque being mostly linear in bias while the out-of-plane torque is quadratic. Most importantly, we find that the spin transfer torque in antiferromagnetic tunnel junctions is much more robust against disorder than in antiferromagnetic metallic spin-valves due to the tunneling nature of spin transport.
\end{abstract}
\maketitle

\section{Introduction}
Intensive research has been achieved in the field of spin transfer torque \cite{Slonczewski,Berger,Ralph_MMM2008} in ferromagnetic materials in the last two decades. Spin torque consists of the transfer of spin angular momentum from a spin-polarized flow of conduction electrons to the local magnetic moments of a ferromagnet. This spin transfer promotes magnetic excitations resulting in magnetization switching \cite{sun1999,katine2000,Chiba_prl2004} or self-sustained precessional motion \cite{Kiselev_Nature2003,Rippard_prl2004}.
The typical device on which spin torque switching is commonly achieved is composed of two ferromagnets separated by a spacer that can be either metallic or insulating. The former is henceforth referred to as a metallic spin-valve, while the latter is called a (ferro)magnetic tunnel junction (F-MTJ). In both devices, the spin torque is dominated by an antidamping component of the form $\tau_{\|}\sim{\bf {m}}\times({\bf p} \times {\bf { m}})$, where ${\bf p}$ is the magnetization direction of the reference layer while ${\bf { m}}$ is the magnetization direction of the free layer. In both cases, the torque is an interfacial process arising from the destructive interference between incoming electron spins with different incidences \cite{Stiles_prb2002,manchon2008}. In the case of F-MTJs, a field-like torque of the form $\tau_{\bot}\sim{\bf { m}}\times{\bf { p}}$ also emerges that can be as large as 10 to 30\% of the in-plane torque \cite{theodonis_prl2006,Xiao_prb2008,manchon2008,wil,Heiliger_prl2008}, as confirmed experimentally \cite{Sankey_Nature_phys2008,Kubota_Nature_phys2008,Oh_Nature_phy_2009}. The bias dependence of these two torque components can be tuned by engineering the junction structural asymmetry \cite{kioussis,manchon2010,wil} or in the presence of interfacial electron-magnon scattering \cite{magnon}. 

A few years ago, the presence of spin transfer torque in metallic antiferromagnetic spin-valves has been predicted theoretically \cite{Nunez_prb2006}. The authors considered a structure composed of two antiferromagnetic layers spaced by a normal metal in analogy with the ferromagnetic spin-valve. Experimentally, the search for current induced torque in antiferromagnetic layers has been carried out by analyzing the alteration that occurs at the level of the exchange bias between ferromagnetic and antiferromagnetic layers in a conventional ferromagnetic spin-valve\cite{Tsoi_prl2007,Urazhdin_prl2007}. However, not much progress has been realized experimentally since then due to the significant difficulty of maintaining sizable torques in these structures, as well as controlling and detecting independently the N\'eel order parameter dynamics. As a matter of fact, it was recently shown \cite{hamed_prb2014,Duine2007} that even a small amount of disorder dramatically reduces the magnitude of the torque. Indeed, in order to preserve large current-driven torques in antiferromagnetic spin-valves, the staggered spin density built up in the reference antiferromagnetic layer has to be transported coherently to the free antiferromagnetic layer. Disorder breaks translational invariance and prevents the coherent transmission of this staggered spin density through the spin-valve. \par

A solution to this issue is to generate local torques, i.e. spin currents and densities that do not need to be transmitted from one part of the device to another. Several strategies have been proposed to date, such as the use of antiferromagnetic domain walls \cite{Hals_prl2011,Swaving2011}, or the exploitation of spin-orbit torques \cite{Zelezny_prl2014}. Another approach is to exploit spin-dependent tunneling transport (see Ref. \onlinecite{Merodio_apl2014}), which is much less sensitive to momentum scattering. Recently, tunneling anisotropic magnetoresistance has been reported in IrMn/MgO junctions\cite{Park_Nature2011,Wang2013}, demonstrating the high quality that can be achieved in such systems. Antiferromagnetic spintronics presents tremendous potential for applications and is now gaining significant momentum\cite{Jungwirth2016}.\par

In the present work, we investigate spin transfer torques in antiferromagnetic tunnel junctions (AF-MTJ). We study the voltage dependence of the spin torque components for a junction composed of symmetric antiferromagnetic electrodes. We finally explore the effect of the disorder on the torque, demonstrating that the torque in AF-MTJ is much more robust against imperfections that in antiferromagnetic metallic spin-valves \cite{hamed_prb2014}. 


\section{Methodology}

\begin{figure}[h!]
  \includegraphics[width=8cm]{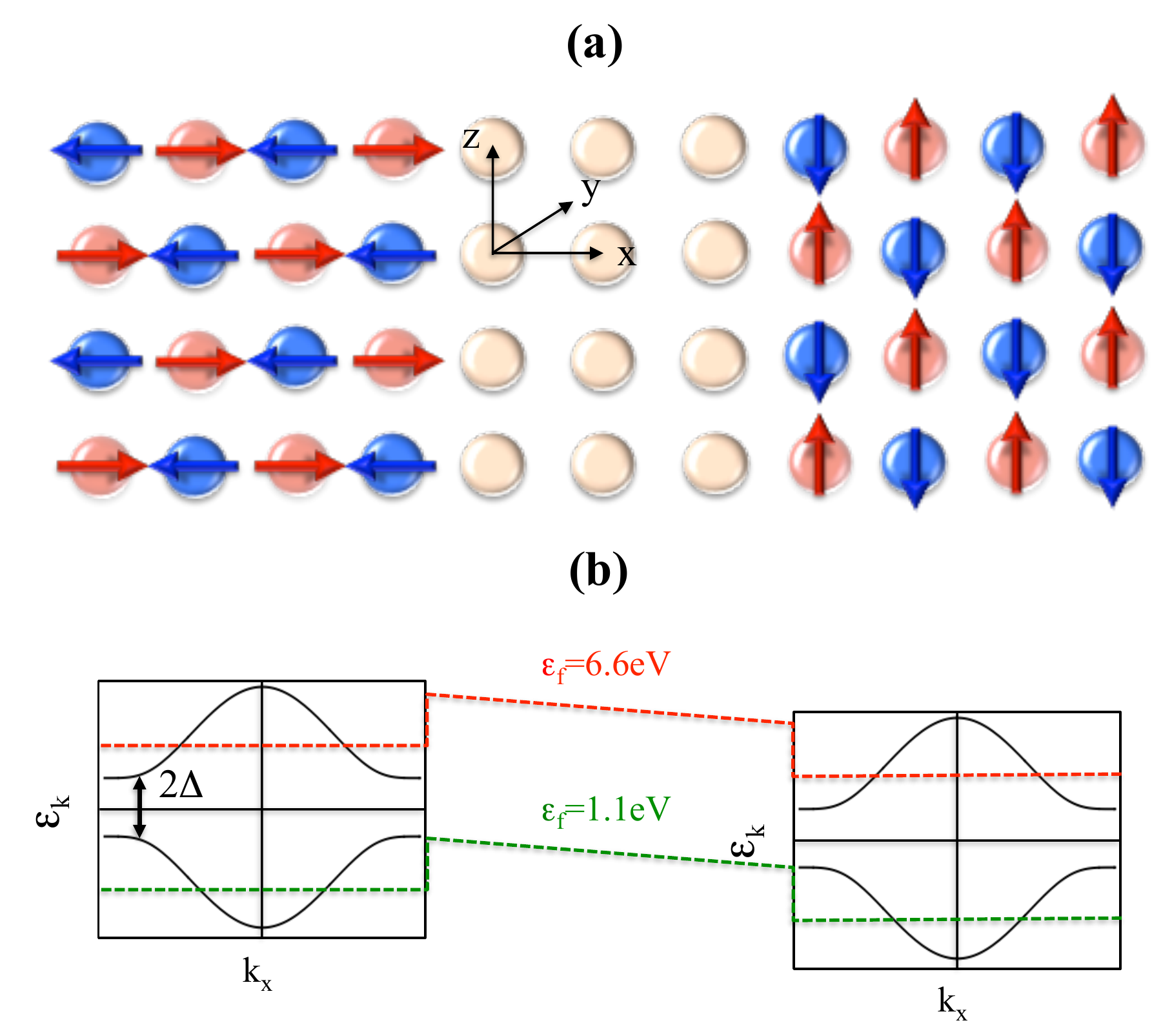}
  \caption{(Color online) (a) Schematics of the AF-MTJ consisting of two semi-infinite antiferromagnets spaced by a tunnel barrier. The red and blue atoms refer to the two antiferromagnetically coupled sublattices. (b) Illustration of the energy spacial profile and its alteration by the applied voltage. The red and green dashed lines represent the potential profile for two specific Fermi energies, $\epsilon_f=6.6$ eV and $\epsilon_f=1.1$ eV, as explained in the text.} \label{fig:introductory}
\end{figure}

The system we consider consists in two semi-infinite antiferromagnetic electrodes spaced by an insulating barrier (see Fig. \ref{fig:introductory}). The two-dimensional antiferromagnets are square lattices in G-type magnetic configuration, i.e. each magnetic moment is surrounded by nearest neighbor moments of opposite direction. This configuration is different from the ones reported in previous theoretical works (Refs.~\onlinecite{Nunez_prb2006,Merodio_apl2014}) in which the authors consider L-type antiferromagnets composed of uncompensated layers with magnetic moments pointing in opposite directions. The width of the junction's layers is $20$ atomic sites while the barrier extends over 3 monolayers. In order to compute the transport properties of this system, we exploit the non-equilibrium Green's function formalism implemented on the tight-binding code KWANT \cite{groth_NewJPhys}, a procedure described in detail in Ref. \onlinecite{hamed_prb2014}. The tight-binding Hamiltonian reads

\begin{eqnarray}\label{eq:Hs}
&&{\hat H}=\sum_{i}\epsilon_{i}{\hat c}_{i}^+ {\hat c}_{i}-\sum_{i,i'}t_{i,i'}{\hat c}_{i}^+ {\hat c}_{i'}+\sum_{i}\Delta_{ex}^i{\hat c}_{i}^+{\bf m}_{i}\cdot{\hat{\bm\sigma}}{\hat c}_{i}.
\end{eqnarray}
The indices $i=(x_{i},y_{i})$ refer to the 2-dimensional coordinates of the sites. $\epsilon_i$ is the on-site energy, $t_{i,i'}=t$ is the hopping parameter between the sites $i$ and $i'$, restricted to nearest neighbors and $\Delta_{ex}^i$ is the exchange energy between the staggered local magnetic moment ${\bf m}_{i}$ on site $i$ and the itinerant electron spin ($\Delta_{ex}^i=\Delta_{ex}$ in the antiferromagnets and $\Delta_{ex}^i=0$ in the barrier). $\hat{\bm \sigma}$ is the vector of spin Pauli matrices where $\hat{}$ denotes a 2$\times$2 matrix in spin space, ${\hat c}_{i}^+$ is the creation operator of an electron on site $i$, such that ${\hat c}_{i}^+=(c_{i\uparrow}^+,c_{i\downarrow}^+)$, where $\uparrow,\downarrow$ refers to the spin projection along the quantization axis.  Compared to the metallic spin-valve studied in Ref. \onlinecite{hamed_prb2014}, the present system comprises a tunnel barrier defined by an onsite energy $\epsilon_{i}^{I,0}=\epsilon_{i}^M+6t$ at zero bias, where $\epsilon_{i}^M$ is the onsite energy of the metallic electrodes at zero bias. The non-equilibrium regime is promoted by applying a voltage $V$ in the range [-0.9t,0.9t] across the junction. The chemical potentials of the two antiferromagnetic electrodes are given by $e\mu^{L(R)}=\epsilon_f\pm eV/2$ and the tunnel barrier onsite energy then reads $\epsilon^I_{i}=\epsilon^{I,0}_{i}+eV[1/2-x_i/(L_B-1)]$, $L_B$ being the barrier length.\par

 The nonequilibrium properties are computed from the lesser Green's function $ {\hat G}^<_{i;i'}(\epsilon)= \sum_l f(\epsilon,\mu_l)\sum_n i\psi^l_{n,i}(\psi^l_{n,i'})^*$ (see Ref. \onlinecite{wimmer_thesis}), $\psi^l_{n,i}$ being the scattering wave functions originating from lead $l$, with a Fermi-Dirac distribution $f(\epsilon,\mu_l)$. In the present work, we calculate the spin torque components from the local spin density ${\bf S}_i$, which reads
\begin{eqnarray}
{\bf S}_{i}&=&\frac{1}{2\pi}\int \text{Tr}_\sigma[\hat{\bm \sigma}\hat{G}^<_{i;i}]d\epsilon.
\end{eqnarray}
The integration runs over the full energy bandwidth up to the chemical potential of the left and right electrodes. The local torque on a particular lattice site reads ${\bm \tau}_{i}=\Delta_{ex}{\bf m}_{i}\times{\bf S}_{i}$.\par

In antiferromagnets composed of collinear sublattices, two types of torques can be defined at the level of the diatomic unit cell: torques arising from uniform spin densities (i.e. when the spin density is equal on the two sublattices), and torques arising from staggered spin densities (i.e. when the spin density is opposite on the two sublattices). In our previous work \cite{hamed_prb2014}, we called these two types of torque "rotating" and "exchange torques", respectively. In systems without translational invariance, such as spin-valves, both types of torques exist in principle and possess components in and out of the (${\bf n},{\bf p}$) plane, where ${\bf p}$ and ${\bf n}$ are the N\'eel order parmaters of the polarizer and analyzer, respectively. At this stage, we want to point out a mistake in the discussion of our previous work, Ref. \onlinecite{hamed_prb2014}. We claimed that the rotating torque (stemming from uniform spin density) is responsible for the N\'eel order parameter switching, while the exchange torque (stemming from staggered spin density) competes with the antiferromagnetic exchange and is therefore ineffective. This claim is incorrect since external magnetic fields (and hence, uniform spin densities) are unable to manipulate the direction of N\'eel order parameters and only result in spin-flop of the antiferromagnet. This confusion arises from the assumption that antiferromagnetic dynamics somewhat ressembles ferromagnetic one, which is clearly untrue since precession about the antiferromagnetic exchange field is the driving force in antiferromagnets (see also discussion in Ref. \onlinecite{Zelezny_prl2014}). Therefore, only staggered spin densities (producing "exchange" torques) are efficient in controlling N\'eel order parameter direction \cite{Jungwirth2016,Gomonay2014}.\par

\section{Results and discussions}
\subsection{Premises}
\begin{figure}[h!]
  \includegraphics[width=6cm]{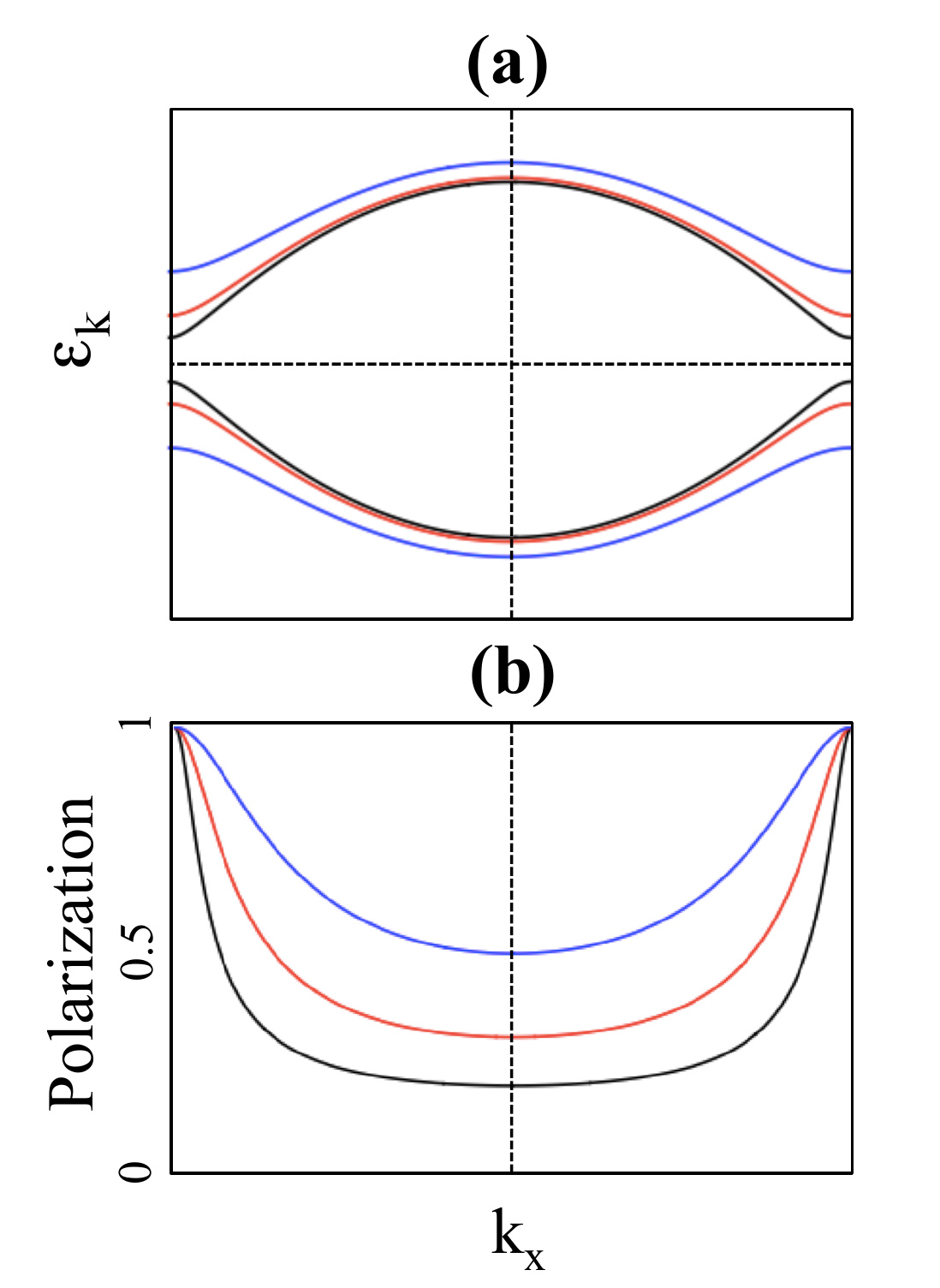}
  \caption{(Color online) (a) band structure and (b) sublattice-resolved spin polarization of a two dimensional G-type antiferromagnet for $\Delta_{ex}=$ 0.5$t$ (black), $t$ (red) and 2$t$ (blue).} \label{fig:bandstructure}
\end{figure}

Before discussing the theoretical results, it is instructive to consider the band structure of a prototypical G-type antiferromagnet. The tight-binding Hamiltonian, Eq. (\ref{eq:Hs}), can be rewritten in the $\{|A\rangle,|B\rangle\}\otimes\{|\uparrow\rangle,|\downarrow\rangle\}$ space, where $A,B$ refer to the antiferromagnetically coupled sublattices. One obtains
\begin{eqnarray}
&&\hat{h}=\gamma_{k}\hat{\tau}_{x}\otimes\hat{1}+\Delta_{ex}\hat{\tau}_{z}\otimes\hat{\sigma}_{z},\\
&&\gamma_{k}=-2t\left(\cos k_{x}a+\cos k_{y}a\right),
\end{eqnarray}
which supports the following eigenstates
\begin{eqnarray}
&&\epsilon_{k}^{s}=s\sqrt{\gamma_{k}^{2}+\Delta_{ex}^{2}},\label{eq:band}\\
&&\psi_{s}^{\sigma}=\frac{1}{\sqrt{2}}\left(\sqrt{1+s\sigma\beta_k}|A\rangle+s\sqrt{1-s\sigma\beta}|B\rangle\right)\otimes|\sigma\rangle,
\end{eqnarray}
and $\beta_k=\frac{\Delta_{ex}}{\sqrt{\gamma_{k}^{2}+\Delta_{ex}^{2}}}$. Here, we chose the N\'eel order parameter to lie along ${\bf z}$ and $\hat{\bm\tau}$, $\hat{\bm\sigma}$ are spin Pauli matrices referring to the $\{|A\rangle,|B\rangle\}$ and $\{|\uparrow\rangle,|\downarrow\rangle\}$ subspaces, respectively. The band structure, Eq. (\ref{eq:band}), is plotted in Fig. \ref{fig:bandstructure}(a) for different values of the exchange energy $\Delta_{ex}$. One obtains the usual gapped electronic structure of antiferromagnets. By increasing the exchange, the band gap increases and the band width is slightly compressed. \par

Although both up and down spin are degenerate, one can define a local spin polarization on each sublattice. In the low band filling limit, $ka\rightarrow0$, the density of state on sublattice $\eta$ ($\eta=1$ corresponds to sublattice A and $\eta=-1$ corresponds to sublattice B) is indeed
\begin{equation}
{\cal N}_{s,\sigma}^\eta\approx\frac{1}{8\pi^2}\left(\frac{2m}{\hbar^2}\right)^{3/2}\frac{\sqrt{4t-\sqrt{\epsilon^2-\Delta_{ex}^2}}}{\sqrt{\epsilon^2-\Delta_{ex}^2}}(-\epsilon+\eta\sigma\Delta_{ex}),
\end{equation}
which produces a polarization 
\begin{equation}
P=({\cal N}_{s,\uparrow}^\eta-{\cal N}_{s,\downarrow}^\eta)/{\cal N}_{s,\uparrow}^\eta+{\cal N}_{s,\downarrow}^\eta=-\eta\Delta_{ex}/\epsilon,
\end{equation}
where $\epsilon$ is measured from the center of the gap. Therefore, the sublattice-resolved polarization is perfect ($P=\pm1$) at the band edges ($\epsilon=\pm\Delta_{ex}$) and decreases to a minimum at the extrema of the bands, as illustrated in Fig. \ref{fig:bandstructure}(b). Since $\epsilon$ remains finite, the minimum polarization never vanishes. It is also remarkable that the polarization is essentially flat, i.e. energy independent, close to $k=0$. This trend is opposite to that of ferromagnets, whose density of state polarization reads $P=(\sqrt{\epsilon+\Delta_{ex}}-\sqrt{\epsilon-\Delta_{ex}})/(\sqrt{\epsilon+\Delta_{ex}}+\sqrt{\epsilon-\Delta_{ex}})$ and decreases when the energy increases away from the bottom of the bands. These features will be essential to understand the robustness of spin torque against disorder in the tunnel barrier, as discussed further below.

\subsection{Spin density profile}

Let us now compute the spin torque components in antiferromagnetic tunnel junctions. In the following, we fix N\'eel order parameter direction of the left antiferromagnet along ${\bf p}={\bf x}$ direction, while the one of the right antiferromagnet points along the ${\bf n}={\bf z}$ direction, as illustrated in Fig. \ref{fig:introductory}. To get better insight about the physics at stake, we will consider two band filling situations: (i) $\epsilon_f=1.1t$, when Fermi energy is located in the middle of the bottom band and (ii) $\epsilon_f=6.6t$, when Fermi energy is located in the middle of the top band (see Fig. \ref{fig:introductory}). In the following, we take $t=1$ eV and the Fermi energy is defined from the bottom of the valence band.\par

\begin{figure}[h!]
  \centering
  \includegraphics[width=9cm]{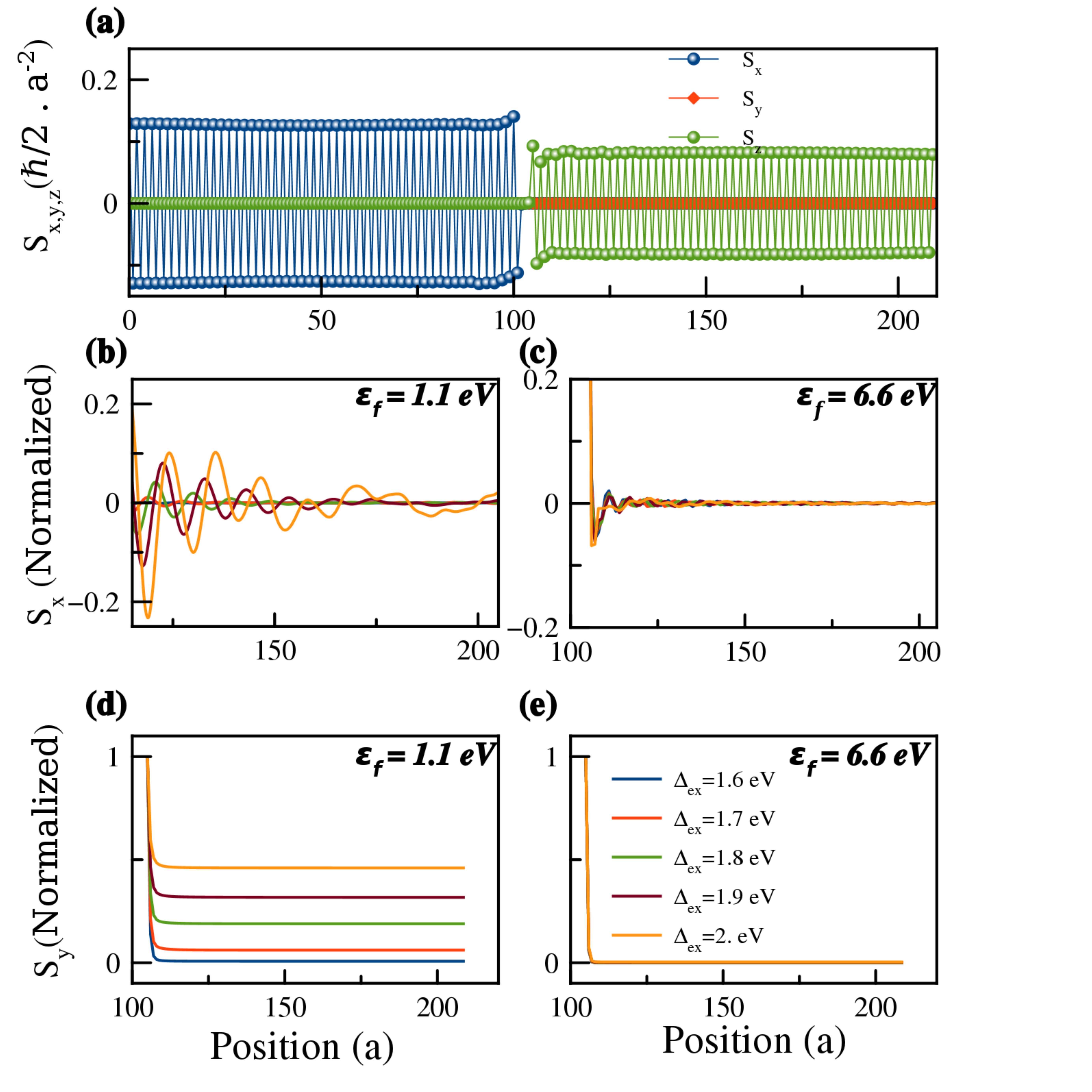}
  \caption{\small{(Color online) (a) Spacial profile of the three components of the spin density $S_{x}$ (blue), $S_{y}$ (red) and $S_{z}$ (green) throughout the junction. The tunnel barrier is located between positions 100 and 103. (b,c) Spatial profile of $S_x$ and (d,e) $S_y$ in the right antiferromagnet, normalized to their magnitude at the right interface, at different exchange energies $\Delta_{ex}$ and for (b,d) $\epsilon_f=1.1$ eV, and (c,e) $\epsilon_f=6.6$ eV. The bias voltage is 0.6 eV.}  \label{fig:SD_ballistic}}
\end{figure}

Upon finite bias voltage, electrons originating from the left antiferromagnetic electrode acquire a staggered spin density along ${\bf x}$ that is injected into the right electrode, as represented by the blue symbols in Fig. \ref{fig:SD_ballistic}(a). In the right, downstream antiferromagnet the itinerant electron spins reorient along the local N\'eel order parameter, i.e. along $ {\bf z}$ direction [see green symbols in Fig. \ref{fig:SD_ballistic}(a)]. During this reorientation, the spin density component transverse to the local N\'eel vector is transferred to the local magnetic moments of the right antiferromagnet. The two spin components transverse to the local N\'eel order parameter of the right layer are reported in Fig. \ref{fig:SD_ballistic}(b,d) and (c,e) at different exchange energies  $\Delta_{ex}$ and for $\epsilon_f=1.1$ eV and $\epsilon_f=6.6$ eV, respectively. For the sake of readability, we report the value of the spin densities normalized to their magnitude at the right interface (actual values of the torque are reported on Fig. \ref{fig:IPOP_ballistic}). \par

The spin density $S_x$ displays a clear oscillatory decay [Fig. \ref{fig:SD_ballistic}(b)], which resembles the behavior observed in F-MTJs (see e.g. Ref. \onlinecite{manchon2008}) but is in sharp contrast though with our previous calculations in metallic spin-valves\cite{hamed_prb2014} where no such decay is observed. The decay increases when increasing the Fermi energy to $\epsilon_f=6.6$ eV [Fig. \ref{fig:SD_ballistic}(c)], and when decreasing the exchange energy $\Delta_{ex}$. We attribute this decay to spin dephasing arising from destructive interference between incoming electrons. Indeed, tunneling involves interference between different bands below the Fermi level. Increasing the Fermi level from 1.1 eV [Fig. \ref{fig:SD_ballistic}(b)] to 6.6 eV [Fig. \ref{fig:SD_ballistic}(c)] increases the number of bands involved in the tunneling process and thereby enhances the spin dephasing. Furthermore, reducing the exchange $\Delta_{ex}$ widens the bandwidth (see Fig. \ref{fig:bandstructure}), which also participates to the enhancement of the destructive interference by allowing more states to tunnel.\par

The $S_y$ component presents a markedly different behavior [Fig. \ref{fig:SD_ballistic}(d)]. It also decays away from the interface, but does not present oscillations. As a matter of fact, while $S_x$ arises from the direct injection of the staggered spin density from the left to the right antiferromagnet, $S_y$ stems from the precession of $S_x$ about the local staggered field. This staggered precession results in a uniform $S_y$ component that presents the same decaying characteristics as $S_x$. Moreover, in the case $\epsilon_f=1.1$ eV $S_y$ reaches a constant value away from the interface, while for  $\epsilon_f=6.6$ eV, it completely vanishes within a few atomic planes [Fig. \ref{fig:SD_ballistic}(e)]. The latter is also a consequence of strong spin dephasing.

\subsection{Voltage dependence}

\begin{figure}[h!]
  \includegraphics[width=8.5cm]{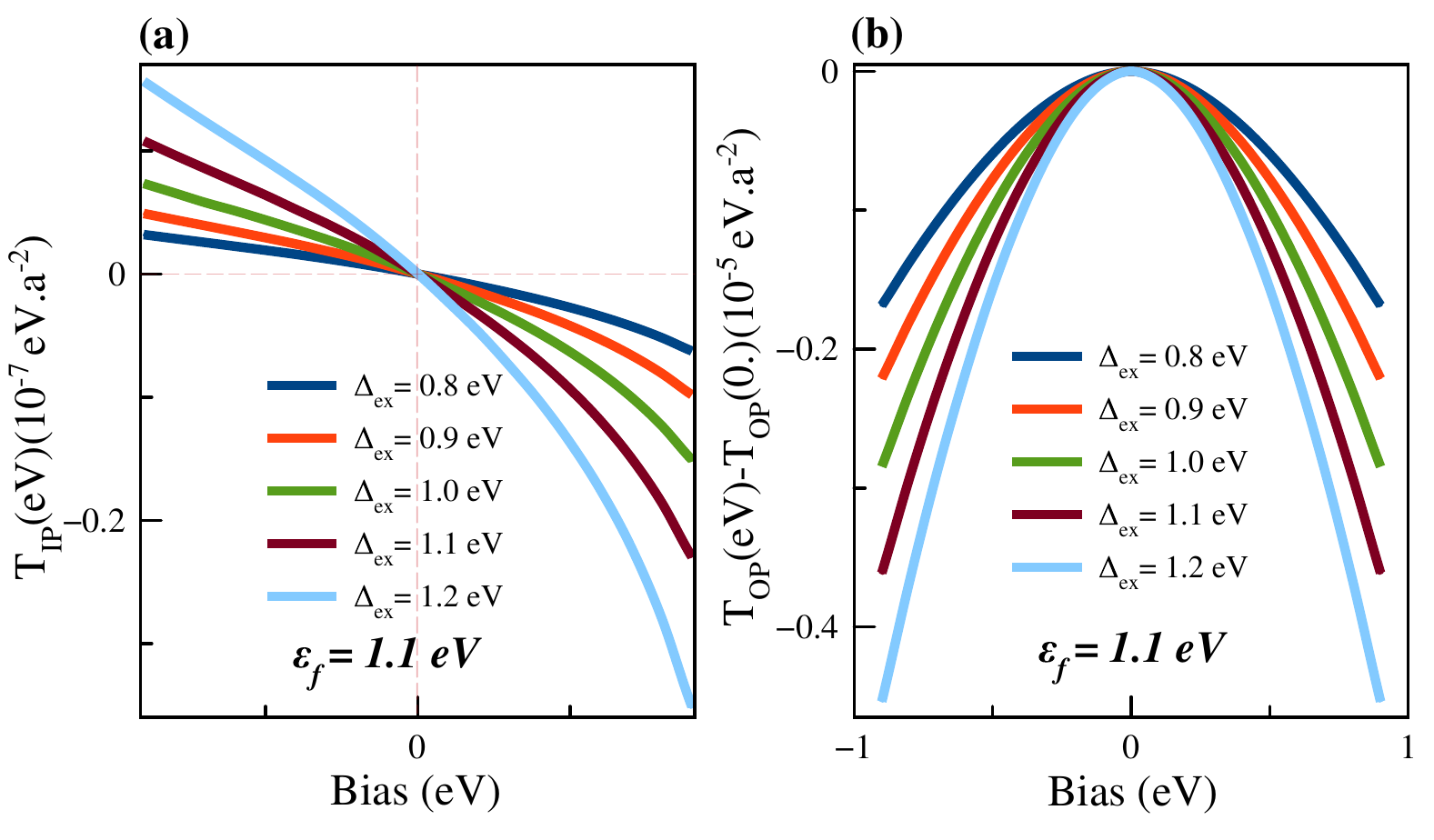}
  \caption{(Color online) Bias dependence of (a) in-plane and (b) out-of-plane torques in AF-MTJ, calculated in the right antiferromagnet for different exchange energy $\Delta_{\rm ex}$ and $\epsilon_f=1.1$ eV.}\label{fig:IPOP_ballistic}
\end{figure}

We now turn our attention towards the bias-dependence of the spin transfer torque. From Fig. \ref{fig:SD_ballistic}, one can anticipate that $S_x$ (which produces the out-of-plane torque, $\sim {\bf n}\times{\bf p}$) provides a dominant staggered density (corresponding to an "exchange" torque in our denomination of Ref. \onlinecite{hamed_prb2014}), while  $S_y$ (which produces the in-plane torque, $\sim {\bf n}\times({\bf p}\times{\bf n})$) provides a dominant uniform density (corresponding to a "coherent" torque in our denomination of Ref. \onlinecite{hamed_prb2014}). As discussed above, only the former enables electrical manipulation of the N\'eel order parameter. To compute the effective torque acting on the layer, we define
\begin{eqnarray}
T_{\rm OP}&=&\Delta_{ex}\sum_{x_i,y_i\in\Omega_R}(S_{x,x_i,y_i}-S_{x,x_i+1,y_i}),\\
T_{\rm IP}&=&\Delta_{ex}\sum_{x_i,y_i\in\Omega_R}(S_{y,x_i,y_i}+S_{y,x_i+1,y_i}),
\end{eqnarray}
where the subscript ${\rm OP}$ (${\rm IP}$) stands for out-of-plane (in-plane) torque component, and $\Omega_R$ is the volume of the right antiferromagnet layer. The unit is given in $t/a^2$, where $t$ is the hopping parameter in eV and $a$ is the lattice parameter (typically 0.4 nm). These torques are reported in Fig. \ref{fig:IPOP_ballistic}(a,b), respectively, for $\epsilon_f = 1.1$ eV as a function of the bias voltage. \par

The bias dependence of these two torques is very similar to what is usually observed in F-MTJs \cite{theodonis_prl2006,manchon2008,Xiao_prb2008,wil}: the in-plane torque displays a bias dependence of the form $T_{\rm IP}= a_1 V + a_2 V^2$, while the out-of-plane torque is mostly quadratic, $T_{\rm OP}=b_0 + b_2 V^2$. Of course, if one introduces asymmetries in the junction, additional linear dependence should appear\cite{wil,manchon2010,kioussis}. Remarkably, the magnitude of the torques reported in Fig. \ref{fig:IPOP_ballistic} is comparable to the one reported for F-MTJs (see, e.g. Ref. \onlinecite{theodonis_prl2006}). We conclude that in a clean AF-MTJ, the current-driven torque efficient to manipulate the N\'eel order parameter direction is an out-of-plane torque that depends on the bias voltage in a quadratic manner. Hence, if one does not break the symmetry of the system, the torque always acts in the same direction, insensitive to the bias polarity.

\subsection{Effect of the disorder}

The impact of spin-independent disorder on spin transport in antiferromagnets is a crucial issue. In general, depending on the growth conditions and techniques, the mean free path is limited by the grain size (i.e. from 5 to 15 nm), which has dramatic impact on spin transport in antiferromagnets\cite{hamed_prb2014}. To implement disorder in our system, we follow the same procedure as in Ref. \onlinecite{hamed_prb2014} and introduce a random onsite potential $\gamma_{i} $ such that $\gamma_{i} \in[-\frac{\Gamma}{2}, \frac{\Gamma}{2}]$, where $\Gamma$ is the disorder strength. At this stage, the computation becomes extremely demanding as both large disorder configurational average together with accurate energy integration are required. In order to ensure the good convergence of our calculation, the Fermi energy is taken at $\epsilon_f=6.6$ eV such that a large number of modes are present in the system, thereby increasing spin dephasing and improving numerical accuracy.

\subsubsection{Disorder in the metallic leads}
\begin{figure}[h!]
  \centering
  \includegraphics[width=8.5cm]{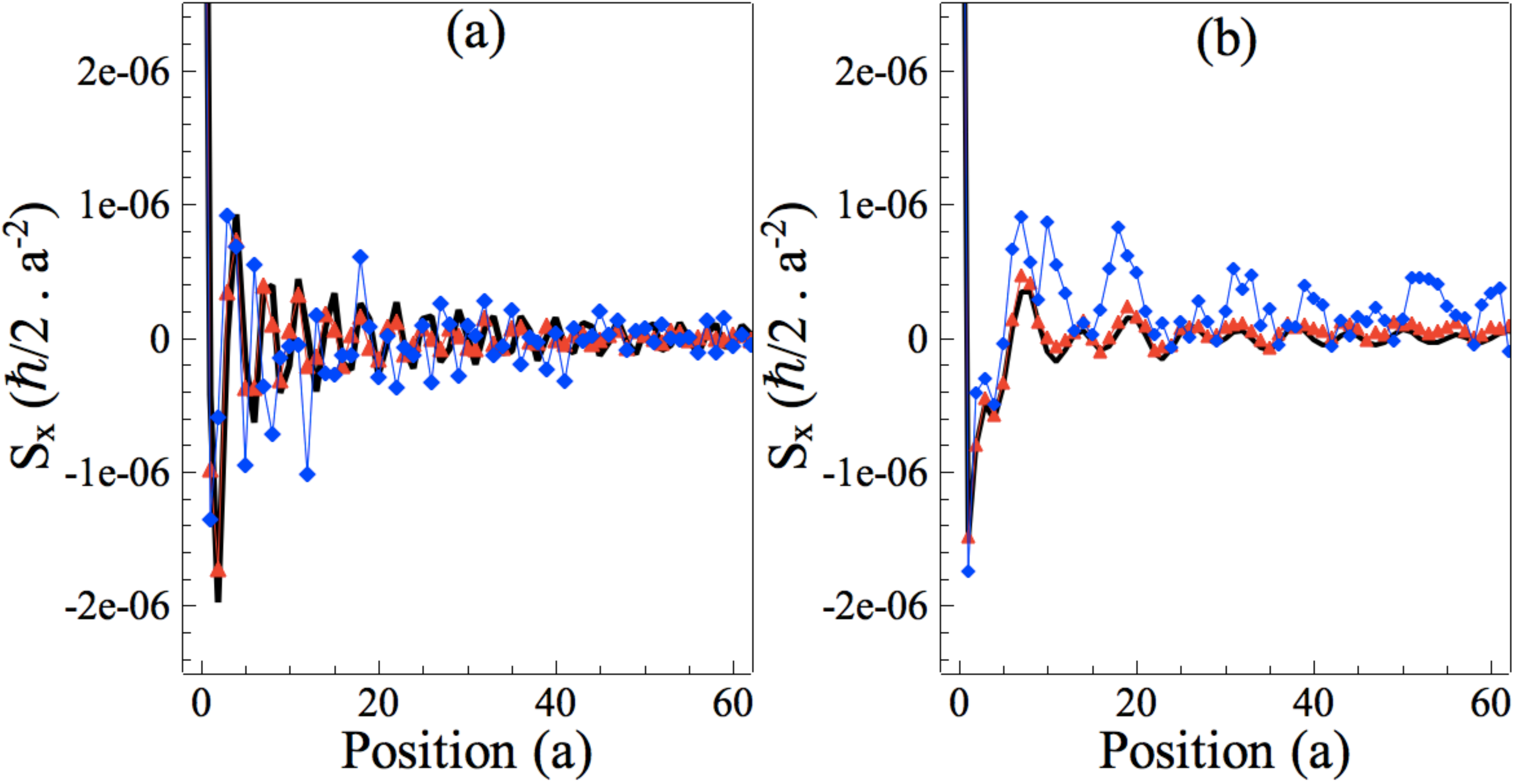}
\caption{(Color online) (a) Spatial profile of $S_x$ in the right antiferromagnet for $\Gamma$=0 (solid line), 0.2 eV (red symbols) and 0.4 eV (blue symbols) for $\Delta_{ex}=1$ eV. (b) Spatial profile of $S_x$ in the right antiferromagnet for $\Gamma$=0 (solid line), 0.1 eV (red symbols) and 0.3 eV (blue symbols) for $\Delta_{ex}=2$ eV. The calculated quantities are averaged over 2000 disorder configurations.}\label{fig:SD_disorderSx}
\end{figure}

\begin{figure}[h!]
  \centering
  \includegraphics[width=8.5cm]{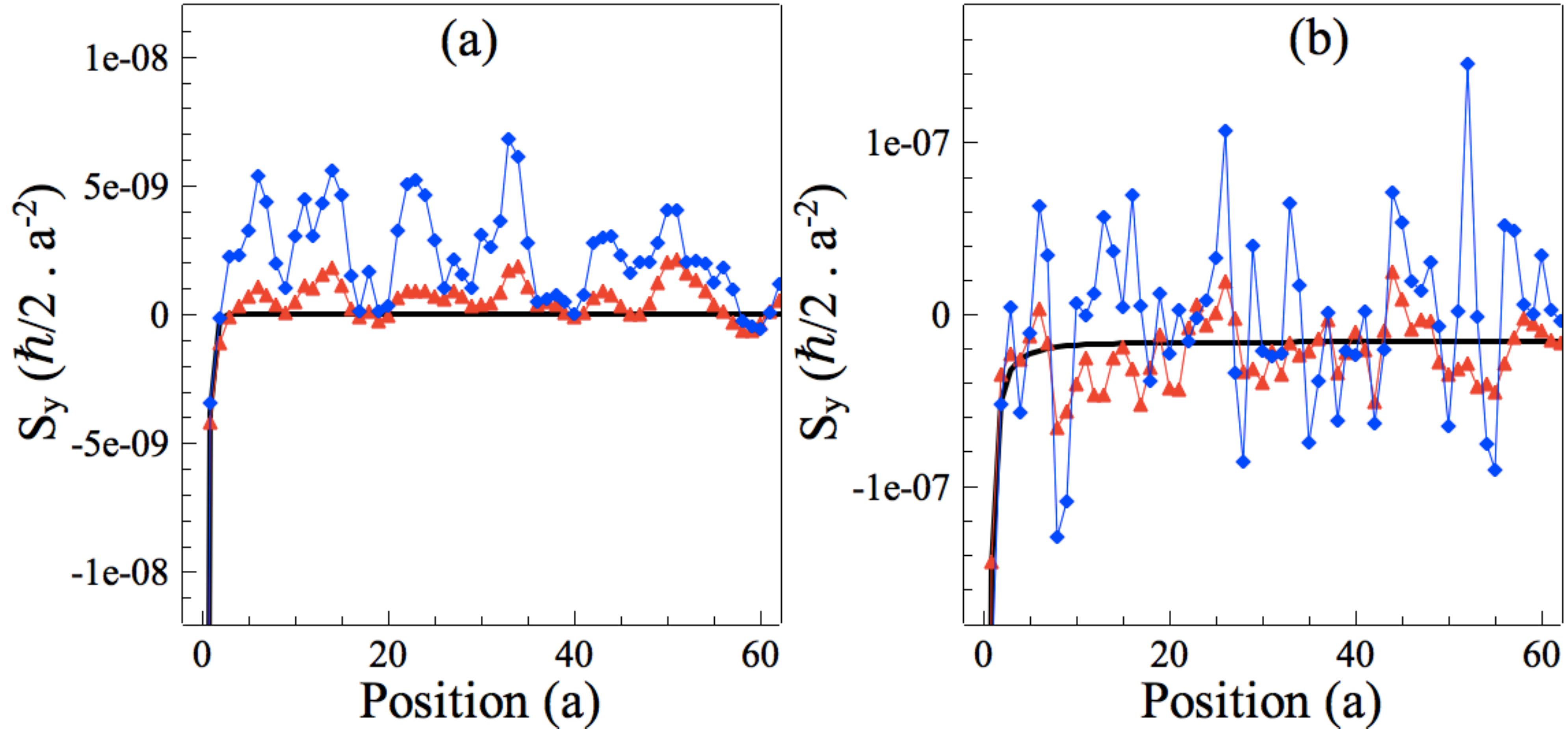}
 \caption{(Color online) (a) Spatial profile of $S_y$ in the right antiferromagnet for $\Gamma$=0 (solid line), 0.2 eV (red symbols) and 0.4 eV (blue symbols) for $\Delta_{ex}=1$ eV. (b) Spatial profile of $S_y$ in the right antiferromagnet for $\Gamma$=0 (solid line), 0.1 eV (red symbols) and 0.3 eV (blue symbols) for $\Delta_{ex}=2$ eV. The calculated quantities are averaged over 2000 disorder configurations.}\label{fig:SD_disorderSy}
\end{figure}

Let us first introduce disorder in the antiferromagnetic leads. Figs. \ref{fig:SD_disorderSx} and \ref{fig:SD_disorderSy} display the spatial profile of $S_{x}$ and $S_{y}$, respectively, for different disorder strengths and exchange energies. The clean regime is also reported for comparison (solid lines in Figs. \ref{fig:SD_disorderSx} and \ref{fig:SD_disorderSy}). Symbols represent the disordered regime with $\Gamma$ ranging from $0.1$ to $0.4$ eV. Since the spin density decreases dramatically within two monolayers from the interface (see Fig. \ref{fig:SD_ballistic}), we focus on the impact of disorder on the oscillatory decay of the spin density in the bulk of the antiferromagnet. For weak disorder (red symbols) in Fig. \ref{fig:SD_disorderSx}(a) and (b), the oscillation of $S_x$ remains weakly affected, while increasing the disorder results in enhanced deviations (blue symbols). However, Fig. \ref{fig:SD_disorderSx}(a) and (b) show that disorder mostly affect the spin density in the bulk of the antiferromagnet, away from the interface. As a result, since the torque mostly occurs at the interface where the spatial decay is stronger, the overall torque magnitude remains only weakly affected by disorder. A similar conclusion can be drawn for $S_{y}$, displayed in  Fig. \ref{fig:SD_disorderSy}. Again the magnitude of $S_{y}$ is mostly affected by bulk disorder, while its value close to the interface remains robust. As a conclusion, the overall impact of disorder on spin torque is much less dramatic than in metallic spin-valve since in AF-MTJ the torque is mainly an interfacial effect.

\subsubsection{Disorder in the tunnel barrier}

\begin{figure}[h!]
  \centering
  \includegraphics[width=9cm]{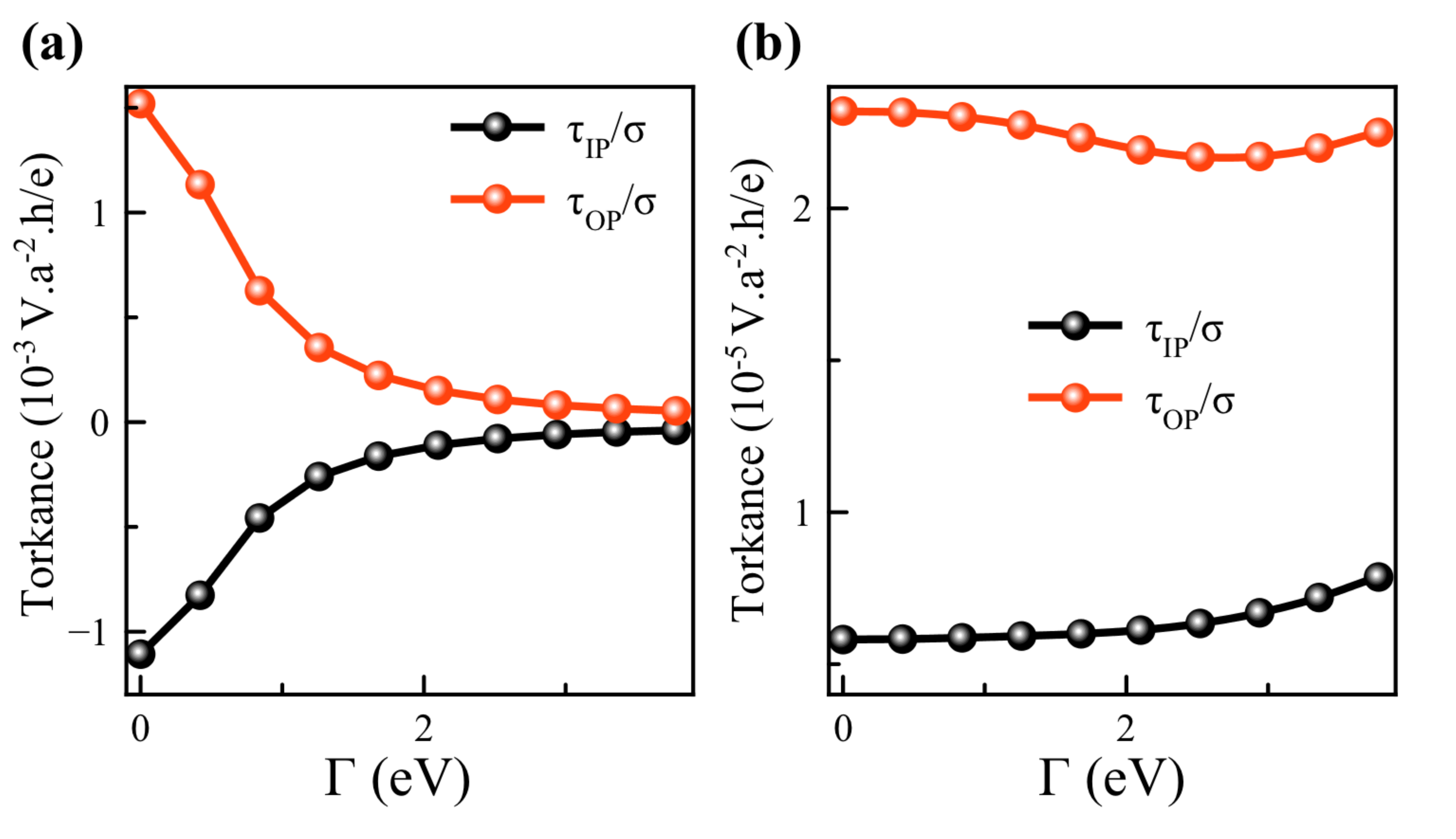}
  \caption{(Color online) Impact of disorder on the torkance components in (a) F-MTJs and (b) AF-MTJs. The black and red symbols refers to the in-plane and out-of-plane components, respectively. The exchange splitting is set to $\Delta_{ex}=2.$ eV, the width of the junctions is $10$ monolayers, the bias voltage is 0.6 eV.}\label{fig:AF-F-disorder}
\end{figure}

The impact of disorder in the tunnel barrier on spin transport has been reported in the case of tunneling magnetoresistance in F-MTJs \cite{Tsymbol_CondMatt2003}. In these structures, the disorder inside the tunnel spacer is detrimental to spin transport properties since a local reduction of the barrier heigh or thickness enhances the tunneling current while reducing its spin polarization. In other words, the presence of disorder in the tunnel barrier introduces hot spots of weakly polarized current that dominate the magnetoresistance signal. Let us now consider the impact of disorder in the barrier on spin torque in F-MTJs. Fig. \ref{fig:AF-F-disorder}(a) shows the torkance - or torque efficiency -, defined as the torque normalized to the conductance, exerted on the right ferromagnetic layer. The torkance (proportional to the interfacial polarization) dramatically decreases in the presence of disorder, which is consistent with the behavior previously observed for the tunneling magnetoresistance\cite{Tsymbol_CondMatt2003}. \par

We now turn our attention towards the effect of disorder on the spin transfer torque in AF-MTJs, reported in Fig. \ref{fig:AF-F-disorder}(b) [the parameters are the same as in Fig. \ref{fig:AF-F-disorder}(a)]. Surprisingly, the torkance remains mostly unaffected by the disorder, in sharp contrast with F-MTJs. This illustrates the major difference between AF-MTJs and F-MTJs: since up and down spins are degenerate [see Fig. \ref{fig:bandstructure} and related discussion], the hot spots introduced by the disorder only results in an enhancement of the tunneling current without altering the sublattice polarization. As a consequence, the spin torque efficiency in AF-MTJs is much more robust against disorder than in F-MTJs.\par


\section{Conclusion}
In the present work we studied the spin transfer torque in AF-MTJs using a real-space tight-binding model. We have shown that, similarly to the case of F-MTJs, the antiferromagnetic torque is interfacial and possesses both in-plane and out-of-plane torques, the former being mostly linear in bias voltage while the latter is quadratic for a symmetric system. However, two main differences have been identified. First of all, since only staggered spin densities are efficient in manipulating N\'eel order parameter, the efficient torque is out-of-plane. Second, because up and down spins are degenerated in antiferromagnets, the torque efficiency in AF-MTJs is much more robust against disorder than in F-MTJs. This shows that AF-MTJs are solid candidates for the realization of spin transfer torque in antiferromagnets.

\acknowledgments
A.M. acknowledges the financial support of the King Abdullah University of Science and Technology (KAUST) through the Office of
Sponsored Research (OSR) [Grant Number OSR-2015-CRG4-2626].

\end{document}